\documentclass[useAMS,usenatbib]{mn2e}
\usepackage{graphicx,amssymb,natbib}

\title{The population in the background of open clusters: Tracer of the Norma-Cygnus arm}

\author[A. K. Pandey et al.]
  {A.~K.~Pandey$^1$\thanks{pandey@aries.ernet.in}, S.~Sharma$^1$ 
 and K.~Ogura$^2$  \\\\
  $^1$Aryabhatta  Research Institute of Observational Sciences, Manora Peak,
       Nainital, 263 129, Uttaranchal, India\\
  $^2$Kokugakuin University, Higashi, Shibuya-ku, Tokyo 150-8440, Japan}

\date{Accepted xxxxxx.  Received xxxxxx.}

\pagerange{\pageref{firstpage}--\pageref{lastpage}}
\pubyear{2006}

\begin{document}

\maketitle

\label{firstpage}

\begin{abstract}

We present colour-magnitude diagrams of open clusters, located in the range $112^\circ < l < 252^\circ$,
manifesting stellar populations in the background of clusters. Some of the populations are found to be
located beyond the Perseus arm and may be the part of Norma-Cygnus (outer) arm. The outer arm seems to be
continued from $l\sim120^\circ$ to $l\sim235^\circ$. The background populations follow the downward
warp of the Galactic plane around $l\sim240^\circ$.

\end{abstract}


\section{Introduction}

\begin{figure*}
\includegraphics[height=6.6in,width=6.6in]{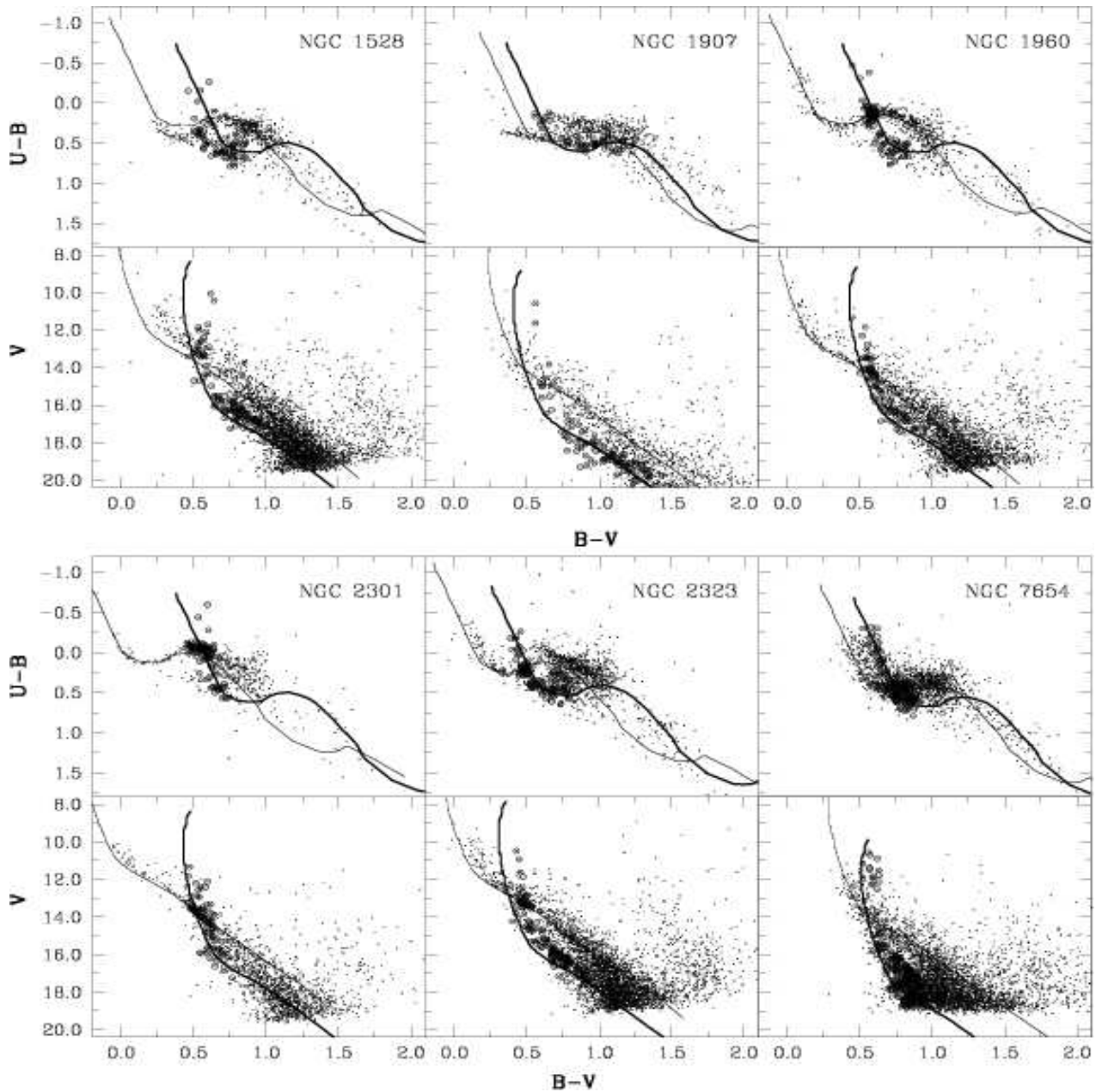}
\caption{Upper and lower panels of the figure show two-colour diagrams and CMDs for the six clusters studied by us. Thin and thick lines represent ZAMS shifted to match the observed sequence of cluster and background population respectively (see  text for details). Open circles represent probable background population members.}
\end{figure*}

\begin{figure*}
\includegraphics[height=6.6in,width=6.6in]{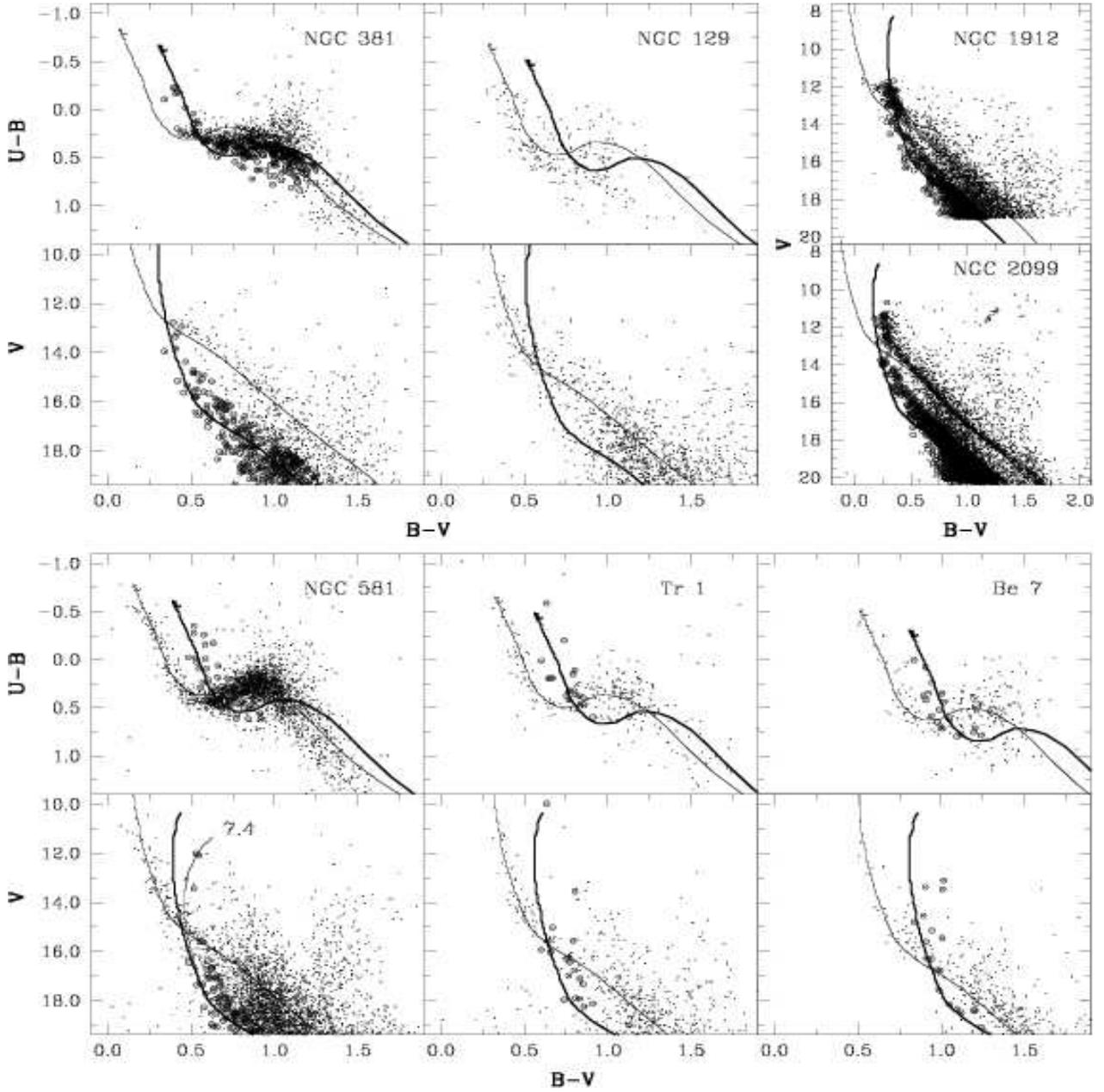}
\caption{Same as Fig. 1 but for data taken from literature. In the case of NGC 129 the background population could not be disentangled. Upper right panel of the figure shows CMDs for the clusters NGC 1912, and NGC 2099 (see  text for details). Isochrone by Bertelli et al. (1994) for log age = 7.4 is also shown in the CMD of cluster NGC 581.}
\end{figure*}

\begin{figure*}
\includegraphics[height=4in,width=7.0in]{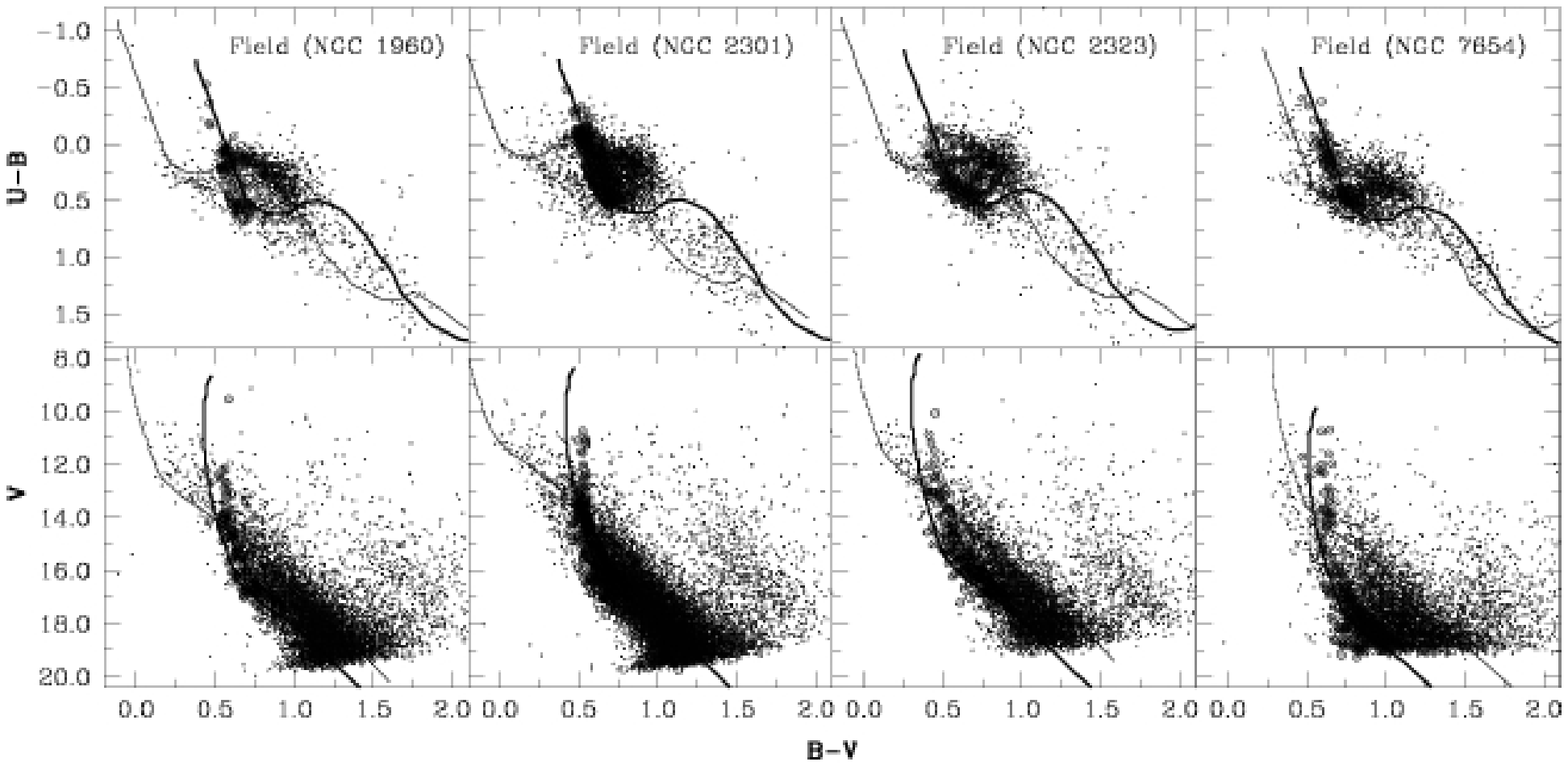}
\caption{Same as Fig. 1 but for nearby field regions to the clusters }
\end{figure*}

The recently discovered ring-like structure (the Monoceros/Canis Major Ring) in the anti-center direction
of the Galaxy (e.g. Newberg et al. 2002, Ibata et al. 2003) could be the consequence of accretion
of dwarf galaxies by the Milky way (cf. Martin et al. 2004, Bellazzini et al. 2006 and references therein).
Helmi et al. (2003) suggested that the ring may be a tidal arc produced by stars stripped away from the 
parent satellite galaxy during a recent interaction. This would produce an asymmetric structure above and below
the disc, limited in Galactic longitude and with a significant velocity gradient in the Galactic longitude
direction. If the ring is symmetric without velocity gradient, it could be produced by ancient minor mergers.

From the analysis of the asymmetries in the population of Galactic M-giant stars, Martin et al. (2004)
have reported detection of the Canis Major dwarf galaxy (CMa) located at 7-8 kpc from the Sun and centered at
$l\sim240^\circ$ and $b\sim -8^\circ$. CMa was identified as strong elliptical-shaped over-density of M-giants by 
the comparison of star counts in Northern and Southern Galactic hemispheres. The $V/B-V$ colour magnitude
diagram (CMD) of a field located at $\simeq4^\circ.2$ from the center of the CMa by Bellazzini et al. (2004)
revealed a population of an intermediate/old ($age\sim4-10$ Gyr) and moderately metal deficient
stellar system located a distance of $d_\odot=8.0\pm1.2$ kpc. These results have also been supported 
by the CMD presented by Martinez-Delgado et al. (2005). A Blue Plume (BP) of possibly young
stars or blue stragglers has also been detected in the CMD. Martin et al. (2005) further reported
existence of CMa on the basis of distance-radial velocity gradient among CMa stars that can 
be explained as the effect of on-going tidal disruption of the stellar system.

\begin{figure}
\includegraphics[height=3.5in,width=3.0in]{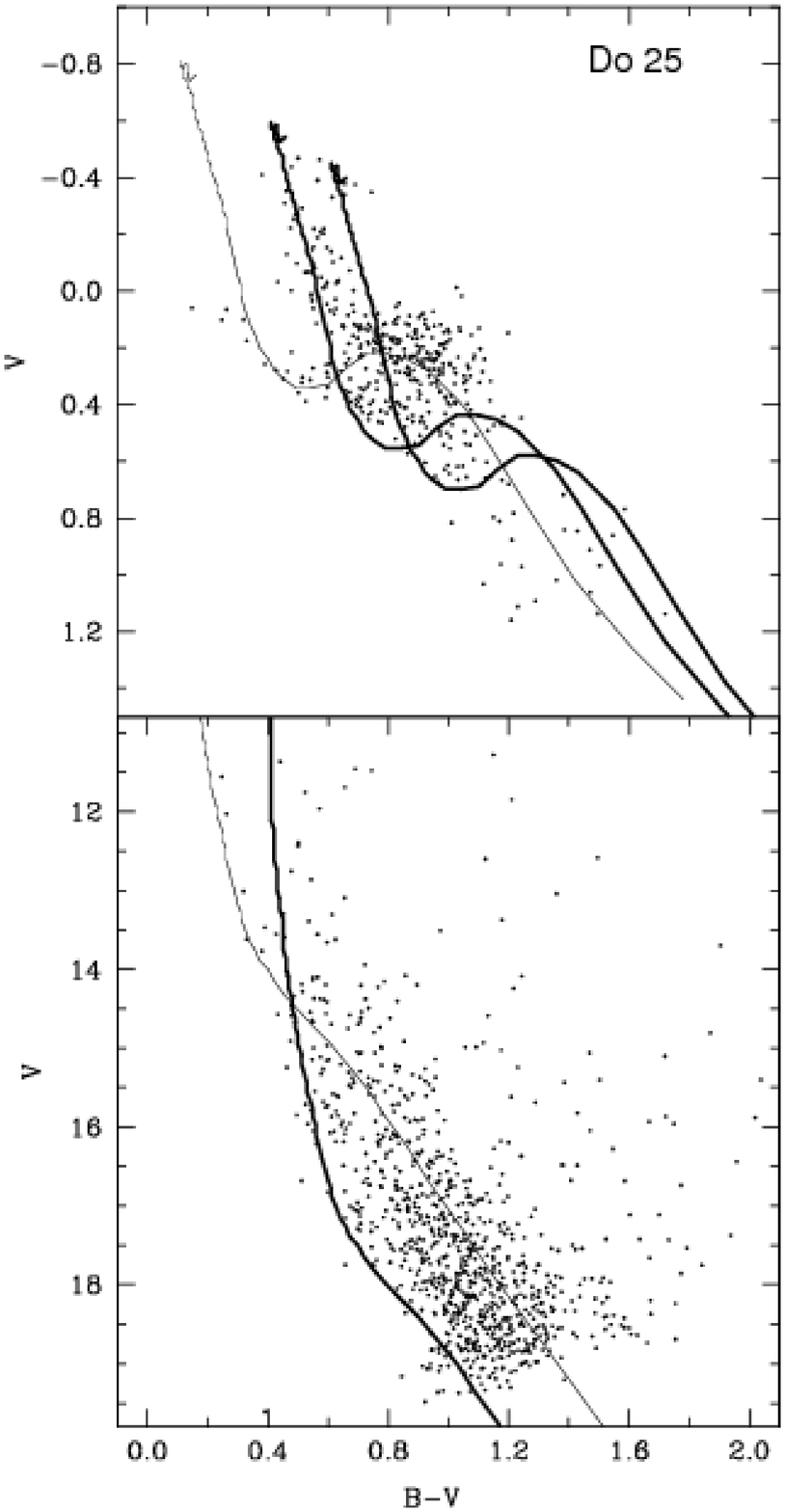}
\caption{Upper and lower panels show TCD and CMD respectively for Do 25 cluster region. Thin and thick lines represent ZAMS shifted to match the observed sequence of foreground and cluster stars respectively. }
\end{figure}

Momany et al. (2004, 2006) suggested that the CMa over-density is an effect of the Galactic
warp. The Southern stellar maximum warp occurring near $l\sim240^\circ$ (for $R_\odot \sim7$ kpc) brings
down the Milky Way mid-plane by $\sim3^\circ$ in this direction.
The regularity and consistency of the stellar, gaseous and dust warp argues strongly against
a recent merger scenario for CMa. They present evidence to conclude that all observed parameters
of CMa are consistent with it being a normal Milky Way outer-disk population.

Carraro et al. (2005) and Moitinho et al. (2006) have also detected BP population (age $\le100$ Myr)
in the background of several open clusters in the third Galactic quadrant. They concluded that BP population
is associated to Norma-Cygnus spiral arm and the over-density towards CMa is simply a projection effect
of looking along the nearby local arm.

The detection of the CMa over-density by Martin et al. (2004) has attracted the attention
towards the third Galactic quadrant of the Milky Way. Young open clusters are one of the 
best tools to study the spiral arms and structure of the Milky Way (Becker \& Fenkart 1970; Pandey et al. 1988, 1990).
Relevant parameters like their distance, age, reddening towards the cluster can be measured more accurately than 
for single stars. We have embarked on a wide field CCD photometric
survey around open clusters using the Kiso Schmidt telescope (Japan), since
extensive studies of the coronal regions of open clusters have not
been carried out so far mainly because of non-availability of photometry in a
large field around open star clusters.
We have undertaken a $CCD$ photometric study of a large area around open
clusters using the $2K \times 2K$ $CCD$ mounted on the 105-cm Schmidt telescope
of the Kiso Observatory which gives $\sim 50'\times 50'$ field.
Results on few open clusters have been reported by Pandey et al. (2001, 2005, 2006) 
and Sharma et al. (2006). We also noticed undesired background contamination
in the CMDs of open clusters located in the second and third galactic quadrant, 
therefore in the light of above discussions it is
considered worthwhile to study the background population towards the target clusters.

\section{Observational data}

We have carried out CCD photometry of several clusters, in the galactic longitude range $112^\circ \le l \le222^\circ$,
during 1999 to 2001 using the 105 cm Schmidt telescope of the Kiso Observatory. The CCD
camera used a SITe 2048$\times$2048 pixel$^2$ TK2048E chip having a pixel size 24 $\mu$m.
At the Schmidt focus (f/3.1) each pixel of CCD corresponds to $1.^{\prime\prime}5$
and the entire chip covers a field of $\sim 50'\times 50'$ on
sky. The read out noise and gain of the CCD are 23.2 $e^-$ and 3.4 $e^-$/ADU respectively.
The analysis presented by Pandey et al. (2001, 2005, 2006) and Sharma et al. (2006) indicates accuracy and
consistency of the data with the previously published work. The data is homogeneous and errors are $\sim0.1$ mag
at $V\sim19-20$ mag. Here we used the data set where $U$ band observations are available because
$U$ band allow an accurate estimation of reddening towards the cluster region.
Details of the clusters are given in Table 1.
 
We have supplemented present data with the data available in the literature. 
Carraro et al. (2005) from their CCD data of eleven open clusters, 
in the longitude range $219^\circ < l < 252^\circ$, have reported a detection of
a young stellar population ($\le 100$ Myr) located at a distance of $d_\odot \sim 7-11$ kpc. They also
estimated reddening towards the BP direction by using $(U-B)/(B-V)$ two colour diagrams. 
The result for eleven clusters namely NGC 2302, NGC 2362, NGC 2367, NGC 2383, NGC 2384,
NGC 2432, NGC 2439, NGC 2533, Be 33, Ru 55 and To 1 have been taken directly from Carraro et al. (2005).
$U,B,V$ CCD photometric data for 5 clusters have been taken from Phelps \& Janes (1994).
The data for another two clusters (NGC 2099 and NGC 2168)
is taken from Kalirai et al. (2001, 2003) and WEBDA. All the relevant parameters of clusters taken from the
literature are given in Table 2.
Thus in the present paper we studied BP towards 26 clusters distributed in the
Galactic longitude range $112^\circ \le l \le 252^\circ$.

\begin{table*}
\centering
\caption{Details of the clusters observed by us.}
\begin{tabular}{@{}rrrcccccrcc@{}}
\hline
\hline
Cluster & $l$ & $b$ & Observed & $E(B-V)$ &$d_\odot$  & Log age & $E(B-V)_{\tiny BP}$& ${d_\odot}_{BP}$& $E(B-V)_{FIRB}$&Reference\\
&          (deg) & (deg)& band           & (mag)    & (kpc) &(yrs)&(mag) & (kpc)&   (mag)&\\   
\hline
NGC 7654   &  112.82  &   +00.43   &$UBVI$  &0.57  &1.4& 8.2 &$0.80\pm0.05$ &$6.4\pm0.5$ &0.95& 1\\ 
NGC 1528   &  152.06  &   +00.26   &$UBVRI$ &0.26  &1.1& 8.6 &$0.72\pm0.10$ &$3.6\pm0.7$ &0.90& 2\\
NGC 1907   &  172.62  &   +00.31   &$UBVRI$ &0.52  &1.8& 8.5 &$0.70\pm0.05$ &$4.6\pm0.5$ &0.97& 3\\
NGC 1912   &  172.25  &   +00.70   &$UBVI$  &0.25  &1.4& 8.5 &$0.60\pm0.05$ &$4.3\pm0.5$ &0.82& 3\\
NGC 1960   &  174.53  &   +01.07   &$UBVRI$ &0.22  &1.3& 7.4 &$0.72\pm0.05$ &$4.2\pm0.5$ &0.95& 2\\
Dolidze 25 &  211.94  &   -01.27   &$UBVI$  &0.70-0.90  &5.8& 6.8 &              &            &2.52& 3\\
NGC 2301   &  212.56  &   +00.28   &$UBVI$  &0.03  &0.9& 8.2 &$0.72\pm0.05$ &$3.6\pm0.5$ &1.15& 2\\
NGC 2323   &  221.67  &   -01.33   &$UBVI$  &0.20  &1.0& 8.0 &$0.60\pm0.05$ &$3.4\pm0.6$ &0.81& 2\\
\hline
\end{tabular}
\footnote{}{1 Pandey et al. (2001); 2 Sharma et al. (2006)(AJ, in press); 3 Pandey et al. (2006, in preparation)}
\end{table*}

\begin{table*}
\centering
\caption{Details of the clusters taken from literature. The results given in columns 5 and 6 are taken
from the reference given in column 11. The results for (*) marked cluster are taken from 
Carraro et al. (2005), whereas result for the cluster Be 17 is also directly taken from literature.
The $E(B-V)_{BP}$ and $d_\odot$ BP for remaining clusters is obtained in the present study.The `@' marked are intermediate populations (see text).}
\begin{tabular}{@{}rrrcccccrcc@{}}
\hline
\hline

Cluster & $l$ & $b$ & Observed  & $E(B-V)$ &$d_\odot$  & Log age & $E(B-V)_{BP}$& ${d_\odot}_{BP}$&$E(B-V)_{FIRB}$&Reference\\
&          (deg) & (deg)&   band & (mag)    & (kpc)      &  (yrs)  & (mag)        & (kpc) &(mag) &\\   
\hline
NGC 129    &  120.27  &   -02.54   &$UBV$   &0.57  &1.7 & 7.7 &$0.80\pm0.08$&$4.8\pm0.6 $&0.89& 1\\
NGC 381    &  124.94  &   -01.22   &$UBV$   &0.36  &1.1 & 9.0 &$0.60\pm0.05$&$4.3\pm0.7 $&0.83& 1\\ 
NGC 581    &  128.02  &   -01.76   &$UBV$   &0.44  &2.7 & 7.3 &$0.68\pm0.05$&$9.5\pm0.5 $&0.59& 1\\
Tr 1       &  128.22  &   -01.14   &$UBV$   &0.61  &2.6 & 7.4 &$0.85\pm0.10$&$7.5\pm0.7 $&0.90& 1\\
Be 7       &  130.14  &   +00.38   &$UBV$   &0.80  &2.6 & 6.6 &$1.10\pm0.05$&$5.2\pm0.5 $&1.27& 1\\
Be 17      &  175.65  &   -03.65   &$BVI$   &0.60  &2.8 & 9.9 &$0.65\pm0.05$&$4.6\pm0.5 $&0.77& 2\\
NGC 2099   &  177.64  &   +03.09   &$BVR$   &0.21  &1.5 & 8.7 &$0.50\pm0.05$&$6.2\pm0.5 $&0.57& 3\\
NGC 2168   &  186.59  &   +02.22   &$BV$    &0.20  &0.9 & 8.3 &$0.50\pm0.05$&$4.9\pm0.5 $&0.60& 3\\
NGC 2302$^*$   &  219.30  &   -03.12   &$UBVRI$ &0.23  &1.5 & 7.1 &$0.70\pm0.05$&$7.5\pm0.5 $&0.83& 4\\
Int 2302$^@$&& &&&&& $0.70\pm0.05$&$4.2\pm0.5 $&0.83&\\
Be 33$^*$  &  225.42  &   -04.62   &$UBVRI$ &0.47  &4.0 & 8.9 &$0.61\pm0.10$&$7.7\pm0.5 $&0.80& 4\\
To 1$^*$   &  232.30  &   -06.31   &$UBVRI$ &0.40  &3.0 & 9.0 &$0.52\pm0.10$&$7.7\pm0.5 $&0.54& 4\\
NGC 2384$^*$   &  235.39  &   -02.39   &$UBVRI$ &0.29  &2.9 & 7.1 &$0.56\pm0.05$&$8.8\pm0.5 $&0.72& 4,5\\
NGC 2383$^*$   &  235.27  &   -02.46   &$UBVRI$ &0.30  &3.4 & 8.1 &$0.56\pm0.05$&$8.8\pm0.4 $&0.73& 4,5\\
NGC 2432$^*$& 235.47  &   +01.78   &$UBVRI$ &0.23  &1.9 & 8.7 &$0.48\pm0.10$&$6.0\pm0.5 $&0.76& 4\\
NGC 2367$^*$& 235.64  &   -03.85   &$UBVRI$ &0.05  &1.4 & 6.7 &$0.62\pm0.05$&$8.5\pm0.5 $&1.07& 4\\
NGC 2362$^*$   &  238.18  &   -05.55   &$UBVRI$ &0.13  &1.4 & 6.7 &$0.40\pm0.15$&$10.8\pm1.2 $&0.53& 4\\
Int 2362$^@$   && & &&&&$0.40\pm0.15$ &$4.8\pm0.5$&0.53&\\
NGC 2439$^*$   &  246.44  &   -04.47   &$UBVRI$ &0.37  &1.3 & 7.0 &$0.47\pm0.10$&$10.9\pm1.1 $&0.58& 4\\
Int 2439$^@$   &&&&&&&$0.47\pm0.10$ &$6.9\pm0.5$&0.58&\\
NGC 2533$^*$& 247.80  &   +01.31   &$UBVRI$ &0.14  &1.7 & 8.8 &$0.48\pm0.05$&$6.5\pm0.3 $&0.67& 4\\
Ru 55$^*$  &  250.69  &   +00.80   &$UBVRI$ &0.45  &4.6 & 7.0 &$0.50\pm0.05$&$7.0\pm0.5 $&0.80& 4\\

\hline
\end{tabular}
\footnote{}{1 Phelps \& Janes (1994); 2 Bragaglia et al. (2006); 3 Kalirai et al. (2001, 2003);
4 Carraro et al. (2005); 5 WEBDA }
\end{table*}

\section{Analysis}

First of all we determined the reddening towards the direction of cluster region using the $(U-B)/(B-V)$
two-colour diagram (TCD). The reddening is obtained by using the slide-fit method and assuming a normal reddening law i.e., $E(U-B)/E(B-V)$ = 0.72. Details of the procedure are given in earlier works (Pandey et al. 2001, 2005). The reddening determination using the $(U-B)/(B-V)$ TCDs is independent of distance of the objects. 
The colour-magnitude diagrams (CMDs) are used to derive the distance to the clusters. Following the traditional method, the fitting was done visually. The estimated random errors in determination of distance modulus of star clusters varies from $\sim$ 0.10 mag to $\sim$ 0.30 mag (see e.g. Kalirai et al. 2001, 2003; Pandey et al. 2001, 2005; Phelps \& Janes 1994; Sagar et al. 2001; Sanner et al. 2001)

The TCDs for six clusters and five clusters obtained from the Kiso Schmidt observations and from data
given by Phelps \& Janes (1994) are shown in Figs. 1 and 2  respectively. A careful inspection of TCDs indicates presence of different populations as reported by Carraro et al. (2005). The reddening for the cluster population is obtained by shifting the unreddened MS along a normal reddening vector $(E(U-B)/E(B-V)=0.72)$ to match with the observed clusters MS. These fits are shown as thin line in Figs 1 and 2.
The unreddened MS is further shifted to match the more reddened observed sequence which is shown by continuous thick line in Figs 1 and 2. This population is referred as blue plume (BP) by Carraro et al. (2005). Then we selected probable BP stars lying along the continuous thick line and the same were plotted as open circles in the TCDs and ($ V, B-V$) CMDs of the cluster region, also shown in lower panels of Figs 1 and 2. 
It can be seen on the TCDs that the spectral type of BP stars in all the CMDs varies from 
early B to late A and F as already indicated by Carraro et al. (2005). However, the 
TCDs are entangled for clusters members and BP stars of late B- and early A-type stars.
After estimating the reddening, the CMDs are used to estimate the distance to the cluster and BP by visual fitting the theoretical ZAMS to the observations. The CMDs for 11 clusters along with visually fitted ZAMS are shown in lower panels of Figs 1 and 2. 
In the case of NGC 129 the background populations could not be disentangled using the TCD and CMD, 
however the TCD and CMD both clearly show the presence of background population. Here we would also 
like to point out that the distance estimation of background population towards the clusters Tr 1 
and Be 7 is based on few points as observations are available for smaller region ($\sim 6.6$ arcmin square). 
However a well defined sequence for the background population can be seen.
Upper right panel of Fig. 2 shows CMDs for the clusters NGC 1912 and NGC 2099 where CCD observations in $\it U$ band are not available, hence $E(B-V)$ is obtained using the photoelectric and/or photographic data. Various parameters obtained from the CMDs are given in Tables 1 and 2. Columns 5,6,7 of the tables give the reddening, distance and age of the clusters respectively. Whereas the results for BP populations are given in columns 8 and 9 of Tables 1 and 2. Column 11 gives the source of the data.
Estimated random errors in determination of distance modulus of BP population is $\sim$ 0.3 mag.
The distance of the most of the BPs obtained in the present work (columns 9, Tables 1 and 2) in the longitude range
$110^\circ < l < 190^\circ$ is $d_\odot \le 6.4$ kpc.
Column 10 gives the reddening estimates towards the cluster direction from the far-infrared background (FIRB)
maps of Schlegel et al. (1998), which are in general higher than the photometric estimates. There are
several studies which have compared interstellar extinction in different region of the Galaxy
with FIRB reddening map and concluded that the FIRB map overestimates the extinction 
(Dutra, Santiago \& Bica 2002; Dutra et al. 2003 a,b; Joshi 2005).

The observations taken through Kiso Schmidt telescope cover an area of about $50^\prime \times 50^\prime$
on the sky, which gives an opportunity to study the field star population located out side the
cluster region. In Fig. 3 we have plotted the TCDs and CMDs for the field region located near
the four clusters. As expected the sequence for cluster stars in the TCDs and CMDs, shown in Fig. 3, is diluted but the prominent sequence for the BPs can be easily noticed just like in Figs 1 and 2. The reddened ZAMS for the same $E(B-V)$ and $(m-M_V)$ as used in Fig. 1 is also shown in Fig. 3. Fig 3 further supports the BPs in the galactic latitude range $110^\circ < l < 222^\circ$.

Here it is worthwhile to mention our preliminary results regarding the cluster Do 25 ($l = 211^\circ.94,
b = -01^\circ.27)$. Fig. 4 shows TCD and $V, B-V$ CMD for the cluster region. 
The TCD indicates a variable reddening, varying from $E(B-V)_{min} = 0.70$ to $E(B-V)_{max} = 0.90$ mag, in the cluster region. The cluster region is strongly affected by the foreground population. The distance and age to the cluster Do 25 are found to be 5.8 $\pm$ 0.5 kpc and 6 Myr. The distance and age of the the cluster are comparable to those of BPs towards the direction of the cluster Do 25.

An inspection of the CMD of three open clusters namely NGC 2302, NGC 2362 and NGC 2439 by Carraro et al. (2005, their figure 1) indicates that there seems to exist a younger population between the cluster and the BP population highlighted by the shaded region in Fig. 5. We shifted the ZAMS (assuming the same $E(B-V)$)
fitted by Carraro et al. (2005) upwards by 1.25 mag, 1.75 mag and 1.0 mag respectively 
in the case of clusters NGC 2302, NGC 2362 and NGC 2439, which indicates a distance $\sim4.2$, 4.8 and 6.9 kpc for the intermediate background population toward the direction of clusters NGC 2302, NGC 2362 and NGC 2439 respectively. The new fitting in the case of NGC 2302 and NGC 2362 is shown in Fig. 5 by thick lines. These populations may be the part of extension of Perseus arm.

\begin{figure}
\includegraphics[height=2in,width=3.3in]{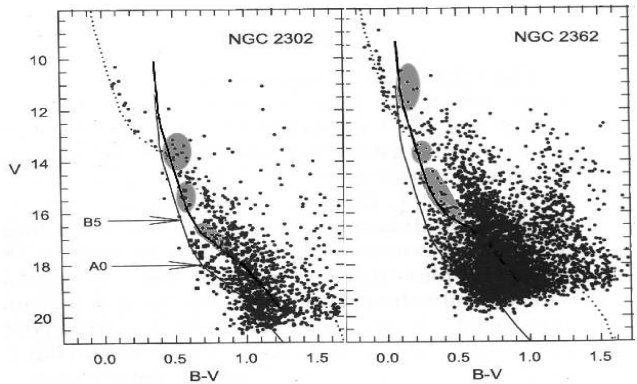}
\caption{Thin line shows the fitting given by Carraro et al. (2005). 
Upward shifted ZAMS, fitted to the stars lying in shaded region, is shown by thick line. For details see text. }
\end{figure}

\begin{figure}
\vbox{
\includegraphics[height=3.1in,width=3.2in]{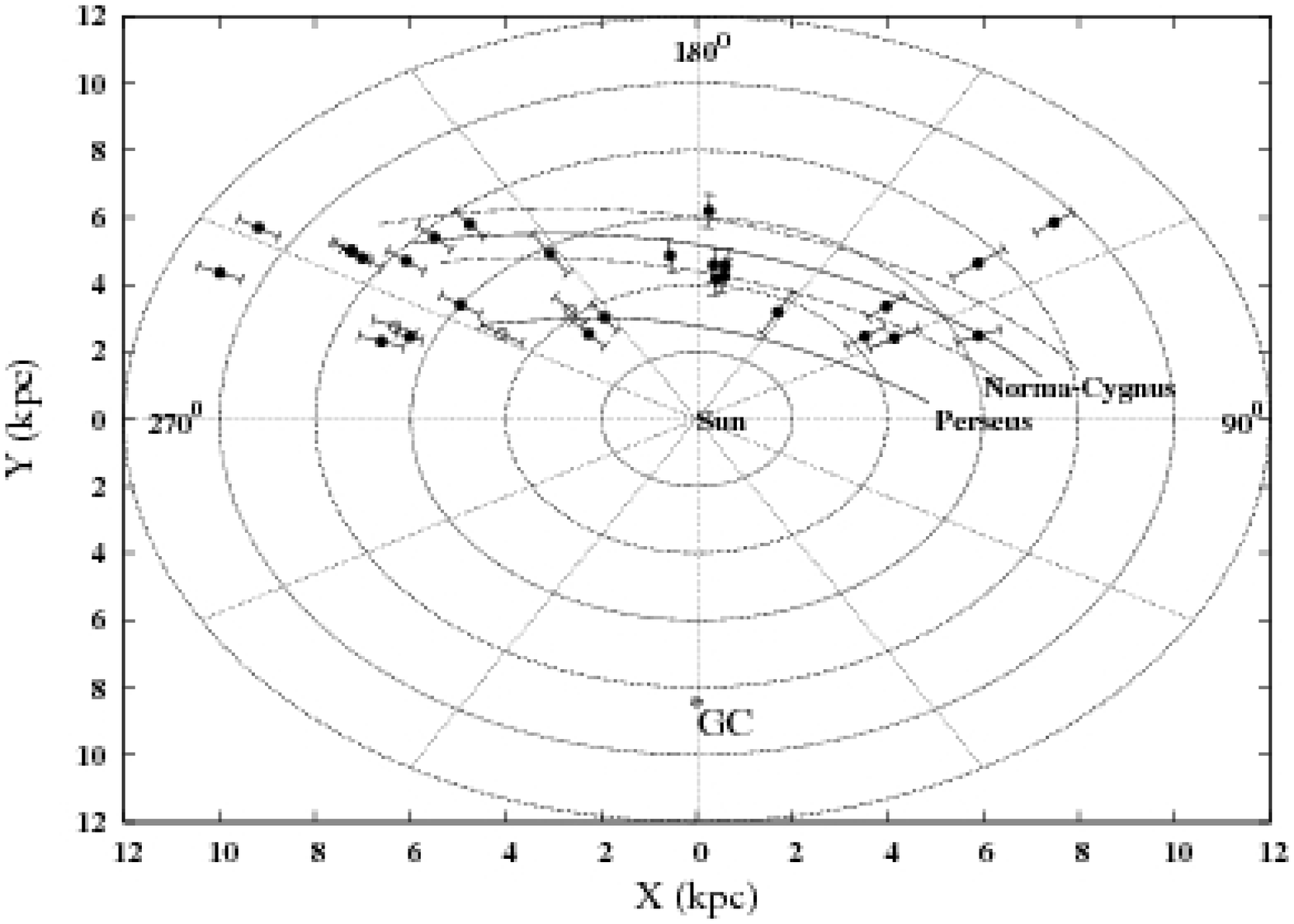}
\includegraphics[height=1.5in,width=3.3in]{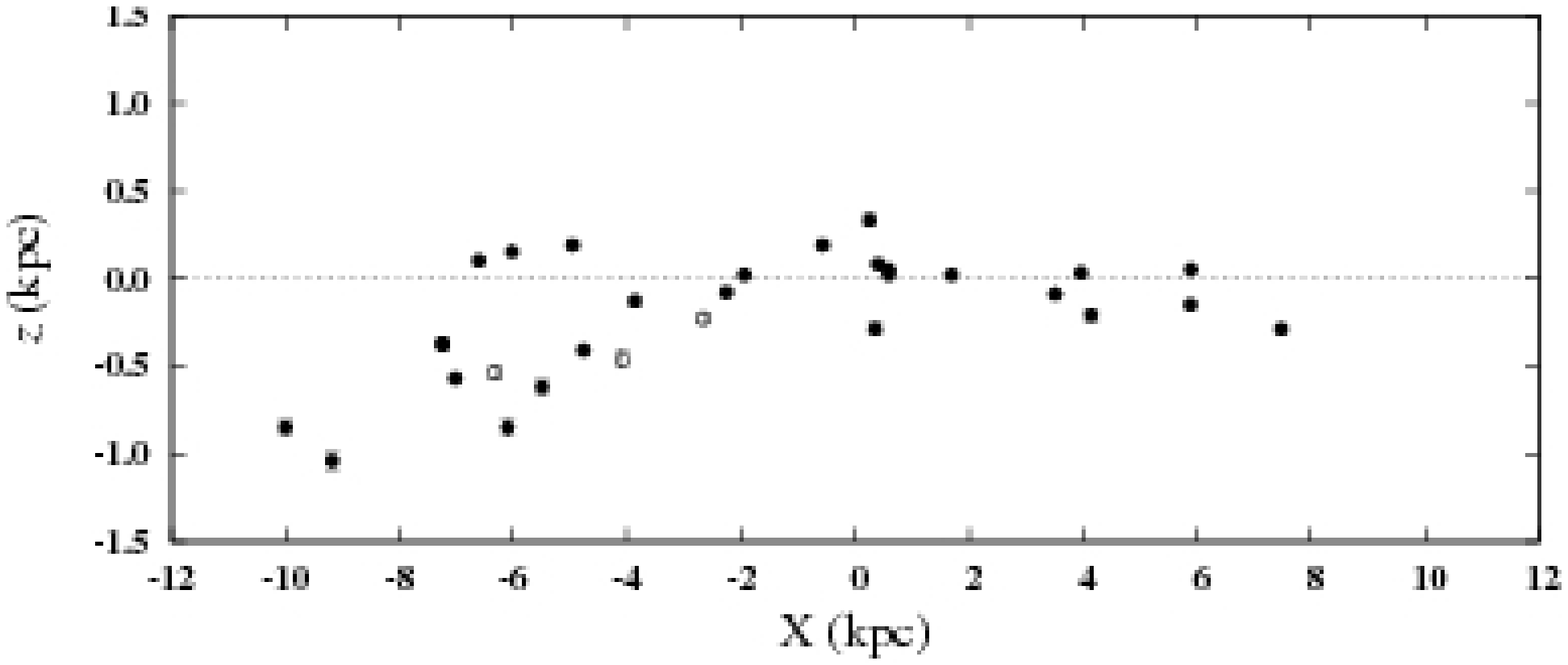}
\includegraphics[height=1.5in,width=3.3in]{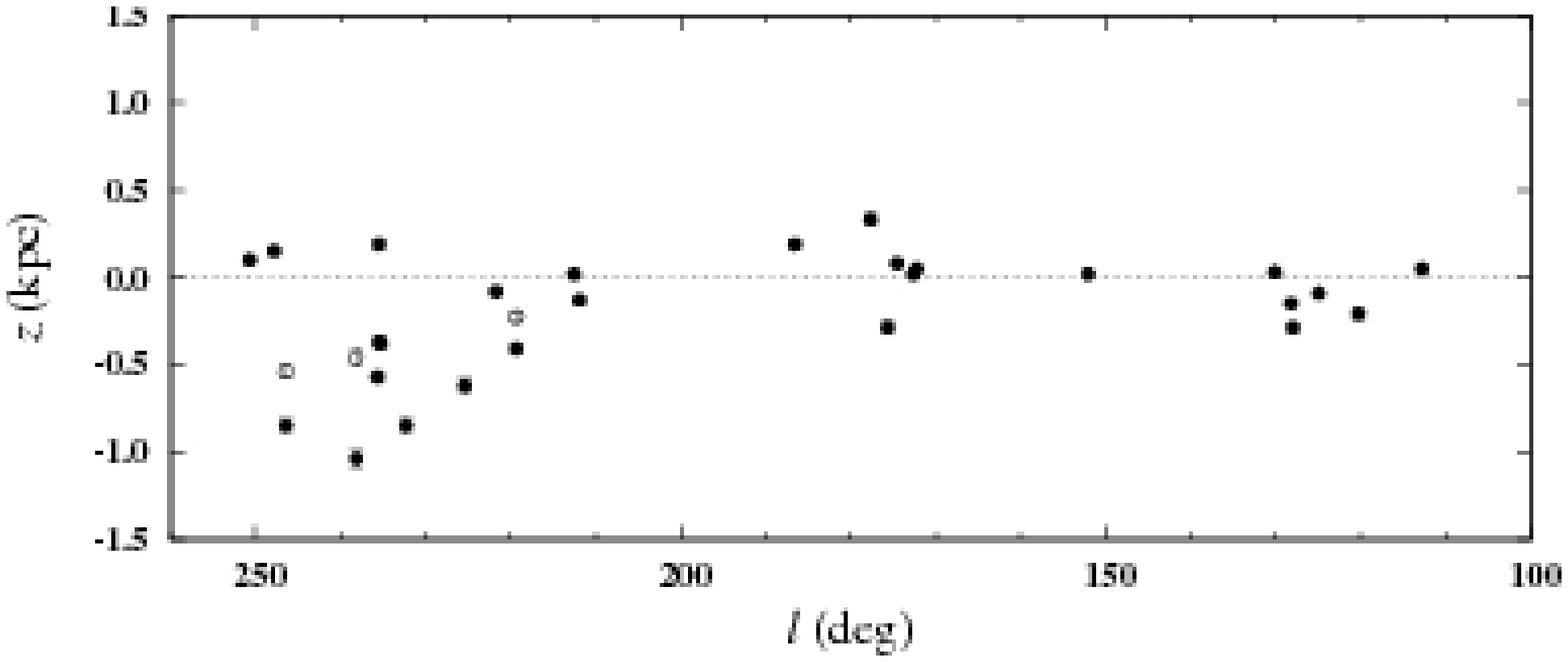}
}
\caption{Upper panel shows the distribution of BPs in the X-Y plane. The continues curves represent Perseus and Norma-Cygnus (outer) arms taken from Russeil (2003). A strip within the thin dashed lines represents a $\sim$1.7 kpc wide outer arm. The galactocentric distance to the Sun is assumed to be 8.5 kpc and the galactic centre (GC) is marked by `*' symbol.  Middle and lower panel show the distribution of BPs in X-Z and $l$-Z plane. Open circles represent intermediate population (see text). }
\end{figure}

\section{Results}

The BP is noticed in the CMDs of all the clusters listed in Tables 1 and 2.
In the case of NGC 2302, NGC 2362, NGC 2439 we noticed some intermediate
background population lying in between the cluster and BPs. 
The distance from the Sun to the BPs is in the range 
$4<d_\odot<9$ kpc and the BPs seem to have a younger 
population of spectral type $\sim B5$ and later. The distribution of BPs
in the longitude range $112^\circ < l < 250^\circ$, listed in Tables 1 and 2 
are plotted in Fig. 6 in X-Y, X-Z and $l$-Z plane of the Galaxy. The two curves shown in the 
upper panel of the figure are location of the outer (Norma-Cygnus) arm and Perseus arm taken from Russeil (2003). 
Momany et al. (2006) have estimated a FWHM $\sim$1.7 kpc for the Norma-Cygnus arm. A strip of about 1.7 kpc wide to represent the outer arm is also shown. The BPs towards the clusters 
NGC 1528, NGC 2301, NGC 2323, NGC 2432, NGC 2533 and Ru 55 indicate that these may be part of the 
Perseus arm, whereas BPs towards the clusters Be 7, Be 17, Be 33, NGC 129
NGC 1907, NGC 1912, NGC 1960, NGC 2099, NGC 2168, NGC 2367, NGC 2383, NGC 2384, NGC 7654, To 1 and cluster Do 25 in the longitude range $112^\circ < l < 250^\circ$ seem
to follow the Norma-Cygnus (outer) arm. Although in our sample we do not have objects in the outer arm in the Galactic longitude range $180^\circ < l < 225^\circ $, however Negueruela and Marco (2003)
have reported that beyond $l=170^\circ$ there is number of open clusters which clearly
delineates the outer arm. They conclude that the existence of structure at d=5-6 kpc over the
$l=175^\circ - 215^\circ$ range seems rather secure.

On the X-Z and $l$-Z projected diagrams, the BPs in the range $112^\circ<l\le220^\circ$
remain close to the formal Galactic plane and the outer arm seems to descend for $l>220^0$
indicating a signature of Galactic warp as discussed by Carraro et al. (2005) and Moitinho et al. (2006). The BP populations towards three clusters NGC 2432, NGC 2533 and Ru 55 that seem to deviate and lie above the galactic plane in $l$ - Z diagram are in fact closer to the Sun and follow the expected extension of the Perseus arm (Carraro et al. 2005). 
The Galactic warp was revealed early from HI data (Kerr 1957), which was further supported
from (1) the $l-z$ distribution
of various spiral arm tracer populations, e.g. open star clusters etc, as a function of longitude
(e.g. Pandey \& Mahra 1987, Pandey et al. 1988, 1990; Djorgovski \& Sosin (1989) and (2) the ratio of star counts in 
Northern and Southern hemisphere as a function of longitude (e.g. Lopez-Corredoira et al. 2002). The 
maximum upward and downward bending occurs at $l\sim60^\circ$ and $l\sim240^\circ$ respectively.

\begin{figure}
\hbox{
\includegraphics[height=1.6in,width=1.6in]{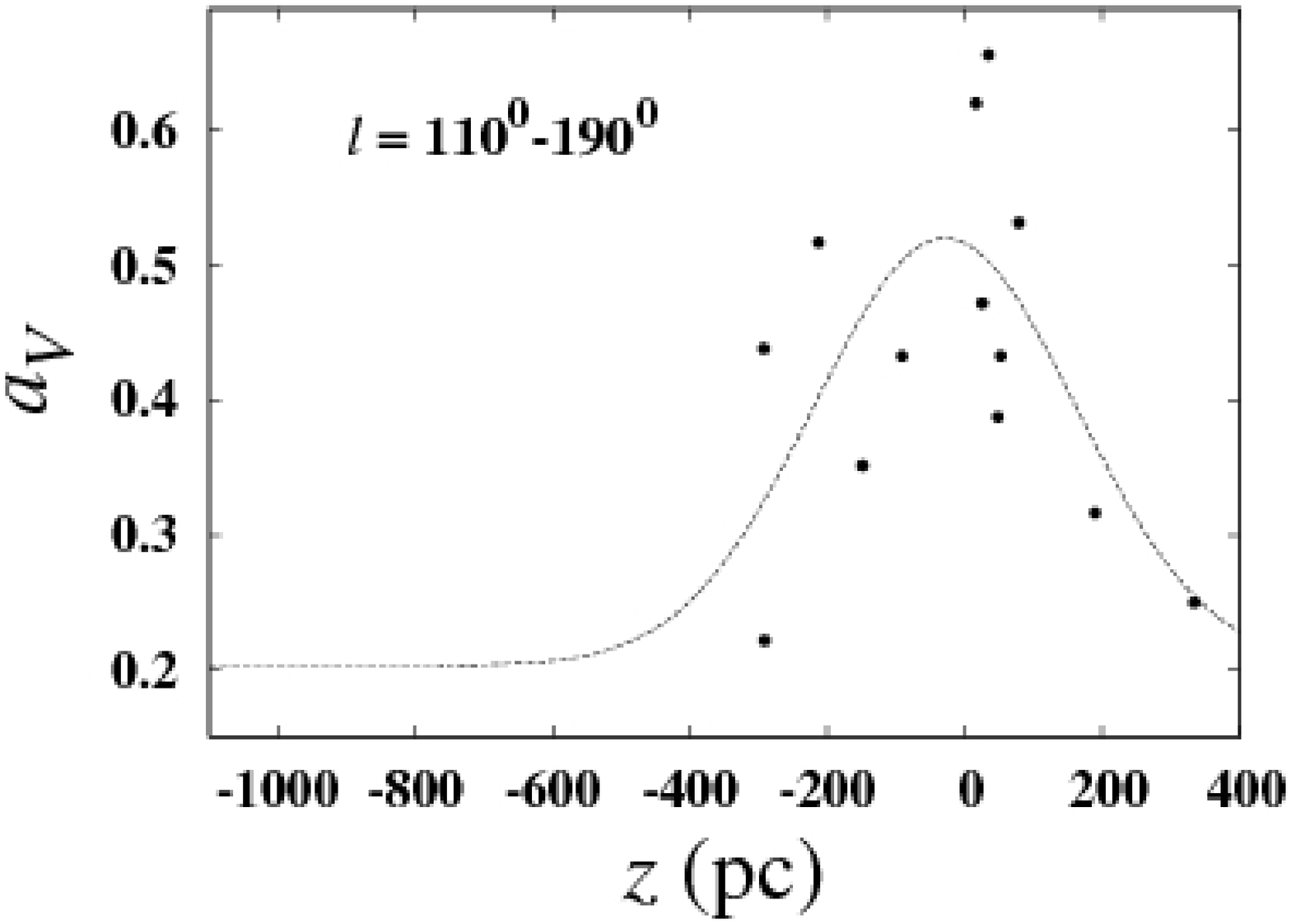}
\includegraphics[height=1.6in,width=1.6in]{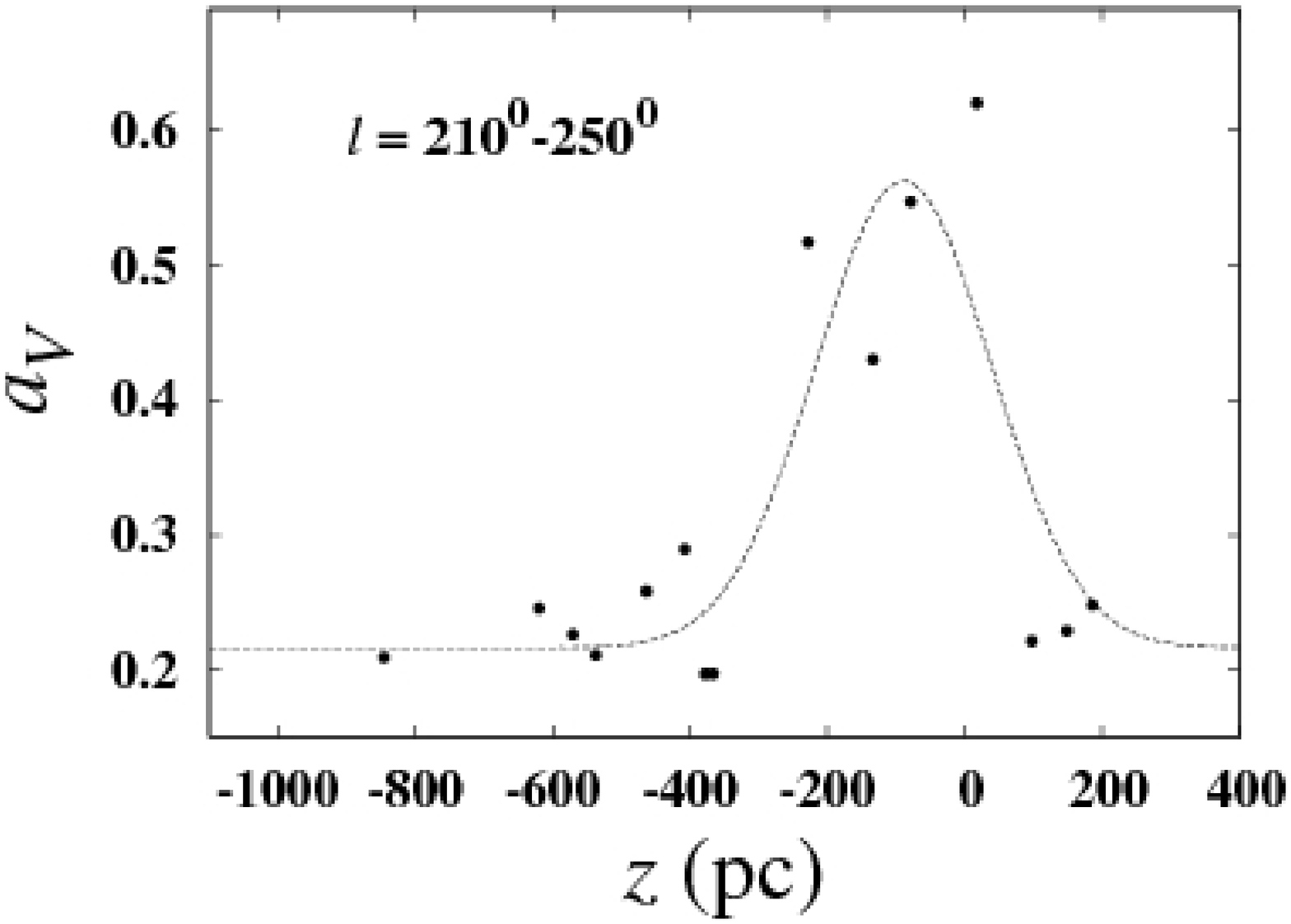}
}
\caption{The distribution of $a_V$ as a function of $z$ towards BPs in the longitude range
$l=110^\circ-190^\circ$ and $l=210^\circ-250^\circ$. }
\end{figure}

To study the distribution of reddening material towards the BPs we plotted interstellar
absorption, $a_V=A_V/d_\odot$ mag/kpc for the Galactic longitude range 
$112^\circ<l<190^\circ$ and $210^\circ < l < 250^\circ$
as a function of $z$ distance in Fig. 7. A Gaussian fit to the distribution yields a peak of $a_V$
at $z\simeq -30\pm 50 $ pc and $z\simeq -90\pm 25$ pc for the range $112^\circ<l<190^\circ$ and $210^\circ<l<250^\circ$
respectively. 

It will be worthwhile to compare the distribution of reddening material towards the BPs with
the distribution of reddening material near the Sun. For this purpose we followed 
the approach described by Pandey \& Mahra (1987) and obtained distribution of $a_V$ as a function
of $z$ in 11 zones of galactic longitude within 3 kpc from the Sun using the catalogue of
open clusters compiled by
Dias et al. (2002). We obtained the value of $z$ for which the absorption is at
maximum ($z_0$) using the Gaussian fit. In Fig. 8 we plot the value of $z_0$ as a function of
longitude. A least-square solution for sinusoidal function manifests that the plane defined by
reddening material shows a maximum upward tilt towards $l\sim60^\circ \pm10^\circ$ or
maximum downward tilt towards $l\sim240^\circ \pm10^\circ$. This result is in agreement with our
previous findings (Pandey \& Mahra, 1987). Fig. 8 indicates that in the longitude range
$110^\circ<l<190^\circ$ (mean $l$ = $150^\circ$), $z_0$ has a mean value of about -25 pc and at 
$l=245^\circ$, the $z_0$ is found to be $\sim -100$ pc. The obtained values of $z_0$ towards above 
mentioned longitudes are in good agreement with the results obtained for the BPs in Fig. 7. 
Here we have to keep in mind that the downward maximum tilt towards $l\sim240^\circ$ found in Fig. 8 
should be considered as a local downward maximum for the population located within few kpc of the Sun. 
Therefore the BPs towards $l\sim240^\circ$  having $d_\odot \le 7$ may be a part
of warped galactic spiral arms. Recently Momany et al. (2006) using 2MASS clump and red giant stars at 
mean heliocentric distances of  $d_\odot$ = 2.8, 7.3 and 16.6 kpc have confirmed southern hemisphere warp 
maximum around $l\sim240^\circ$. Therefore the BPs around $l\sim240^\circ$ and having $d_\odot \ge 8$ 
kpc may also be a part of warped outer galactic arm.

\begin{figure}
\includegraphics[height=2in,width=3.3in]{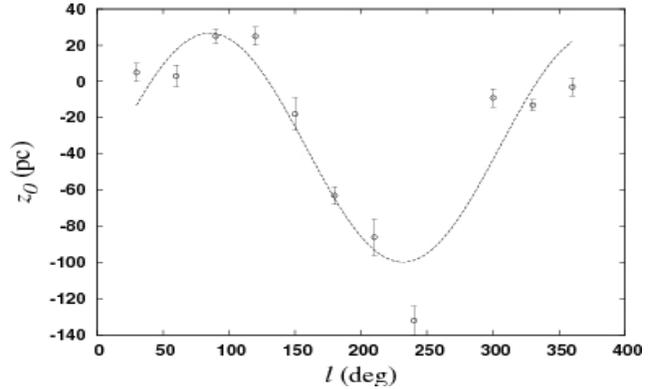}
\caption{Variation of $z_0$ within 3 kpc of the Sun as a function of Galactic longitude. }
\end{figure}

The over-density of stars in the direction of CMa has been explained as a nearby galaxy is being 
cannibalized by the Milky way (Martin et al. 2004, Bellazzini et al. 2004, 2006) A wide sequence
behind the MS of cluster NGC 2477 ($l=253.^\circ6, b=-5.^\circ8$), which is near to the center of CMa
($l\sim240^\circ$, $b\sim-8^\circ$), is attributed to the CMa system (Bellazzini et al. 2004).
However Carraro et al. (2005) pointed out that the same type of populations can also be seen
in the CMD of NGC 2168 ($l=186.^\circ6$, $b=+2.^\circ2$) which is far away from the CMa. Here it is
further interesting to mention that CMD of the cluster NGC 7654 ( $l=112^\circ.8$, $b=+0^\circ.43$),
given by Pandey et al. (2001),
also show a similar type of sequence behind the cluster MS which is found to be situated in
the outer arm at a distance of $\sim6.4$ kpc. A comparison of the CMDs of above mentioned three
clusters is shown in Fig. 9, which clearly indicates a similarity between the CMDs of the clusters.
Here it is interesting to point out that the BP towards two clusters at $l\sim130^\circ$, namely
NGC 581 and Tr 1 are found to be located between $d_\odot \sim 7.5$ to 9.5 kpc. The
work by Russeil (2003) also indicates some population behind the outer arm at $l\sim150^\circ - 225^\circ$.

\begin{figure}
\includegraphics[height=6.6in,width=3.3in]{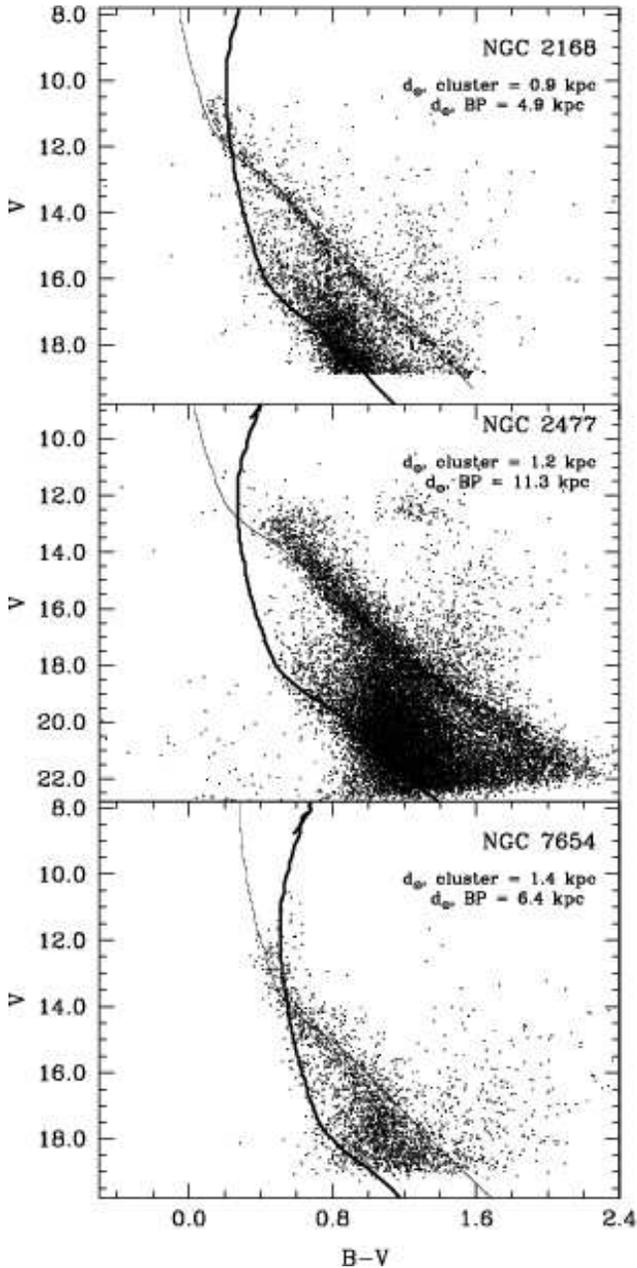}
\caption{CMDs for the clusters NGC 2168, NGC 2477 and NGC 7654. Thin and thick lines shows the visually fitted ZAMS for the cluster and BP respectively. }
\end{figure}

To conclude, the CMDs of the open clusters and nearby field regions used in the present analysis show a significant 
stellar population in background of the clusters, which may be part of the outer spiral arm
as well as of Perseus arm. The structure of Norma-Cygnus (outer) arm 
seems to be continued from $l\sim120^\circ$ to  $l\sim235^\circ$.

Here it is worthwhile to mention that the detection of BP population over a large range of longitudes is not against the dwarf galaxy hypothesis. All the existing models of the disruption of CMa predicts that the remnant and its tidal stream should embrace the whole galactic disc (see e.g., Martin et al. 2005; Pennarrubia et al. 2005). 

\section*{Acknowledgements}

We acknowledge the support given by DST (India) and JSPS (Japan) to carry out the wide field CCD 
photometry around open clusters. We are also thankful to the Kiso observatory for allotting
time at the Kiso Schmidt telescope. Authors are thankful to Dr. M. Bellazzini for useful comments that improved the scientific contents of the paper.

\bsp

\label{lastpage}


\bigskip
\bigskip
\bigskip

\begin{thebibliography}{}
	
\bibitem{}Becker, W., \& Fenkart, R., 1970, in IAU Symp. 38, The Spiral Structure of the Galaxy, ed. W. Becker \& G. Contopoulos (Dordrecht: Reidel), 205
\bibitem{}Bertelli G., Bressan A, Chiosi C., Fagotto F., \& Nasi E., 1994, A\&AS 106, 275
\bibitem{}Bellazzini M., Ibata R., Monaco L., Martin N., Irwin M. J., \& Lewis G. F., 2004, MNRAS, 354, 1263
\bibitem{}Bellazzini M., Ibata R., Martin N., Lewis G. F., Conn B., \& Irwin M. J., 2006, MNRAS, 366, 865
\bibitem{}Bragaglia A., Tosi M., Andreuzzi G., \& Marconi G., 2006, MNRAS, 368, 1971
\bibitem{}Carraro G., Vázquez R. A., Moitinho A., \& Baume G., 2005, ApJ, 630L, 153
\bibitem{}Djorgovski S., \& Sosin C., 1989, ApJ, 341, 13
\bibitem{}Dias W. S., Alessi B. S., Moithinho A., Lepine J. R. D., 2002, A\&A, 389, 871
\bibitem{}Dutra C. M., Santiago B. X., Bica E., 2002, A\&A, 381, 219
\bibitem{}Dutra C. M., Ahumada A. V., Claria J. J., Bica E., Barbuy B., 2003a, A\&A, 408, 287
\bibitem{}Dutra C. M., Santiago B. X., Bica E., Barbuy B., 2003b, MNRAS, 338, 253
\bibitem{}Helmi A., Navarro J. F., Meza A., Steinmetz M., \& Eke V. R., 2003, ApJ, 592L, 25
\bibitem{}Ibata R. A., Irwin M. J., Lewis G. F., Ferguson A. M. N., \& Tanvir N., 2003, MNRAS, 340, 21
\bibitem{}Joshi, Y. C., 2005, MNRAS, 362, 1259
\bibitem{}Kalirai J.S., Ventura P., Richer H.B., Fahlman G.G., Durrell P.R., D'Antona F., \& Marconi G., 2001, AJ, 122, 3239
\bibitem{}Kalirai J.S., Fahlman G.G., Richer H.B., \& Ventura P., 2003, AJ, 126, 1402
\bibitem{}Kerr F. J., 1957, AJ, 62, 93
\bibitem{}Lopez-Corredoira M., Cabrera-Lavers A., Garzón F., Hammersley P. L., 2002, A\&A, 394, 883
\bibitem{}Martin N. F., Ibata R. A., Bellazzini M., Irwin M. J., Lewis G. F., \& Dehnen W., 2004, MNRAS, 348, 12
\bibitem{}Martin N. F., Ibata R. A., Conn B. C., Lewis G. F., Bellazzini M., \& Irwin M. J., 2005, MNRAS, 362, 906	
\bibitem{}Martinez-Delgado D, Butler D. J., Rix H., Franco Y. I., Penarrubia J., Alfaro E. J., \& Dinescu D. I., 2005, ApJ, 633, 205
\bibitem{}Moitinho A., Vázquez R. A., Carraro G., Baume G., Giorgi E. E., \& Lyra W., 2006, MNRAS, 368, 77
\bibitem{}Momany Y., Zaggia S. R., Bonifacio P., Piotto G., De Angeli F., Bedin L. R., \& Carraro G., 2004, A\&A, 421, 29
\bibitem{}Momany Y., Zaggia S., Gilmore G., Piotto G., Carraro G., Bedin L. R., \& de Angeli F., 2006, A\&A, 451, 515	
\bibitem{}Negueruela I., \& Marco A., 2003, A\&A, 406, 119
\bibitem{}Newberg H. J. et al., 2002, ApJ, 569, 245
\bibitem{}Pandey A. K., \& Mahra H. S., 1987, MNRAS, 226, 635	
\bibitem{}Pandey A. K., Bhatt B. C., \& Mahra H. S., 1988, A\&A, 189, 66
\bibitem{}Pandey A. K., Bhatt, B. C., \& Mahra, H. S., 1990, A\&A, 234, 128
\bibitem{}Pandey A. K., Nilakshi, Ogura K., Sagar R., \& Tarusawa K., 2001, A\&A, 374, 504
\bibitem{}Pandey A. K., Upadhyay K., Ogura K., Sagar R., Mohan V., Mito H., Bhatt H.C., \& Bhatt B.C., 2005, MNRAS, 358, 1290
\bibitem{}Pandey A. K., Upadhyay K., Sharma S., Ogura, K., Sandhu T. S., \& Mito H., 2006, submitted
\bibitem{}Pennarrubia, J. et al., 2005, ApJ, 626, 128
\bibitem{}Phelps R. L., \& Janes, K. A. 1994, ApJS, 90, 31	
\bibitem{}Russeil D., 2003, A\&A, 397, 133
\bibitem{}Sagar, R., Munari, U., \& de Boer, K. S., 2001, MNRAS, 327, 23
\bibitem{}Sanner, J., Brunzendorf, J., Will, J. -M., \& Geffert, M., 2001, A\&A, 369, 511
\bibitem{}Schlegel, D. J., Finkbeiner, D.P., \& Davis, M., 1998, ApJ, 500, 525
\bibitem{}Sharma S., Pandey A. K., Ogura K., Mito H., Tarusawa K., \&  Sagar R., 2006, AJ, in press

\end{thebibliography}
\end{document}